\def\ba{\begin{eqnarray}}
\def\ea{\end{eqnarray}}
\def\lb{\label}

\def\bi{\bibitem}

\documentclass[final,3p,times,12pt]{elsarticle}
\usepackage{graphicx,color,amsmath,amssymb}
\usepackage{bm}
\usepackage[colorlinks=true, pdfstartview=FitV, linkcolor=red, citecolor=blue, urlcolor=blue]{hyperref}

\advance\voffset -0.2in
\biboptions{sort&compress}

\begin{document}

\begin{frontmatter}
\title{Multi-particle correlations in transverse momenta from\\statistical clusters}

\author[uj]{Andrzej Bialas}
\ead{bialas@th.if.uj.edu.pl}

\author[agh]{Adam Bzdak}
\ead{bzdak@fis.agh.edu.pl}

\address[uj]{Jagellonian University, M. Smoluchowski Institute of Physics, 30-348 
Krak\'ow, Poland}

\address[agh]{AGH University of Science and Technology, Faculty of Physics and Applied Computer Science, 
30-059 Krak\'ow, Poland}

\begin{abstract}
We evaluate $n$-particle ($n=2,3,4,5$) transverse momentum correlations for pions and kaons following from the decay of statistical clusters. These correlation functions could provide strong constraints on a possible existence of thermal clusters in the process of particle production.
\end{abstract}

\end{frontmatter}

\bigskip

{\bf 1.} Observation of the positive short-range correlations between particles created in high-energy hadron collisions \cite{cor} is naturally interpreted as the consequence of production of particles in "clusters" \cite{abbb,hhlb}. This effect was not unexpected, as the strong interaction between  particles may easily lead to such clustering. The nature of these clusters and the actual mechanism of their formation remains, however, the subject of controversy even now. 

In the present paper we continue the discussion \cite{Bialas:2015pla,Bialas:2015oua} of the consequences of the idea invoked in the statistical cluster model \cite{becattini1,bclus}, where clusters are treated as objects created by  strong interaction in the state of local equilibrium and thus decaying into observed particles according to the Boltzmann distribution, characterized by the inverse cluster temperature $\beta =1/T$. For a cluster moving with the four-velocity $u^{\mu }$ we have
\begin{equation}
dN_{1}(p;u)\sim e^{-\beta p_{\mu }u^{\mu }}d^{2}p_{\perp }dy,
\end{equation}%
where $p_{\perp }$ and $y$ are the transverse momentum and rapidity of the final particle.\footnote{We checked that the Bose-Einstein distribution does not change significantly our results. There are only small corrections (of the order of $10-20$\%) for pions.}
This can be rewritten as  
\begin{equation}
e^{-\beta p_{\mu }u^{\mu }}=e^{-\beta \gamma _{\perp }m_{\perp }\cosh
(y-Y)+\beta p_{\perp }u_{\perp }\cos (\phi _{u}-\phi )},
\end{equation}%
where $u_{\perp }$, $Y$ and $\phi _{u}$ are the cluster transverse four-velocity,
rapidity and azimuthal angle, and $\phi $ is the azimuthal angle of the
produced particle. Finally $m_{\perp}^{2}=m^2 + p_{\perp}^2$, with $m$ being the produced particle mass, and the transverse gamma factor $\gamma _{\perp}^{2}=1+u_{\perp}^2$.

The distribution of the cluster 4-velocity was studied in \cite{Bialas:2015pla} where it was observed that if $\gamma_\perp$  is distributed according to the power law $\sim 1/\gamma_{\perp }^{\kappa }$, the single particle transverse momentum distribution
\begin{equation}
\frac{dN_{1}(p_{\perp })}{dp_{\perp }^{2}}\sim \int_{1}^{\infty }\frac{%
d\gamma _{\perp }}{\gamma _{\perp }^{\kappa }}K_{0}(\beta \gamma _{\perp
}m_{\perp })I_{0}\left( \beta u_\perp p_{\perp }\right)  \lb{ik}
,
\end{equation}%
with $I_{0}$ and $K_{0}$ denoting the modified Bessel functions of the first and the second kind,
follows to a good accuracy the Tsallis distribution \cite{Tsallis:1987eu}, which is known to describe very well the $p_{\perp}$ particle spectra \cite{ww1,Wong:2015mba,AzCl}. Furthermore, it was  shown in
\cite{Bialas:2015oua} that $T=140$ MeV and $\kappa =5$ provides a good description of
pions and kaons in a broad range of $p_{\perp }$. This gives a possible estimate of $T$ and $\kappa$. However, as pointed in Ref. \cite{Becattini:2001fg} the single particle momentum distribution could be distorted by, e.g., secondary particles from resonance decays. We checked that even a substantially  higher temperature (for example $T=160$ MeV) does not modify our results significantly.

In our previous work  \cite{Bialas:2015oua}, we evaluated some of the two-particle correlations resulting from the the production and decay of the uncorrelated  statistical clusters. They present a rather stringent test of the model and can be confronted with experiment. 

Measurements of the two-particle correlations alone, however,  may not be sufficient to distinguish the idea of statistical clusters from  the production of standard  hadronic resonances, representing another possible interpretation of the observed two-particle correlations. It is thus interesting  to 
evaluate and measure also  multi-particle correlations, as they are strongly suppressed in resonance decays while there seems to be no obvious mechanism of such suppression in the decay of statistical clusters.

In this paper we discuss $n$-particle correlations in transverse momenta. It turns out 
that $C_n(p_{1\perp},...,p_{n\perp })$ can be written in a simple analytical form, which can be directly
compared with experiment.
 

{\bf 2.}  If clusters are uncorrelated, the genuine $n$-particle correlation function ($n$-particle cumulant) is
given by the $n$-particle distribution from decay of a single statistical
cluster which reads%
\begin{eqnarray}
\frac{dN_{n}^{(1\,\mathrm{cluster})}}{dy_{1}d^{2}p_{1\perp
}\cdots dy_{n}d^{2}p_{n\perp }} &\sim &\int_{1}^{\infty }\gamma _{\perp
}^{-\kappa }d\gamma _{\perp }\int_{0}^{2\pi }d\phi _{u}\int dYG(Y)\times  \\
&&\times \prod\nolimits_{k=1}^{n}e^{-\beta \gamma _{\perp }m_{k\perp }\cosh
(y_{k}-Y)+\beta p_{k\perp }u_{\perp }\cos (\phi _{u}-\phi _{k})},  \notag
\end{eqnarray}%
where $G(Y)$ is the rapidity distribution of clusters (which should roughly
follow the rapidity distribution of the produced particles) and $u_{\perp
}^{2}=\gamma _{\perp }^{2}-1$.

In this formula the conservation laws are not included. If clusters are objects with well-defined energy, momentum and other quantum numbers, this may lead to error which scales  as inverse of the number of particles in the cluster. However, it is not obvious to us that clusters are isolated from their environment. 
It is actually plausible that they exchange energy, momentum and particles
with their neighbours and thus our canonical description may be an effective
replacement of the exact calculation based on microcanonical ensemble with (unknown) mass, momentum and charge fluctuations.

Performing straightforward integration over $\phi _{k}$ (from $0$ to $2\pi $), 
$y_{k}$ and $\phi _{u}$ we obtain
\begin{equation}
C_{n}(p_{1\perp },...,p_{n\perp })\sim \int_{1}^{\infty }\frac{d\gamma
_{\perp }}{\gamma _{\perp }^{\kappa }}A\left( \beta \gamma _{\perp
}m_{1\perp },...,\beta \gamma _{\perp }m_{n\perp }\right)
\prod\nolimits_{k=1}^{n}I_{0}\left( \beta u_{\perp }p_{k\perp }\right) ,
\label{Cn}
\end{equation}%
where%
\begin{equation}
A\left( \beta \gamma _{\perp }m_{1\perp },...,\beta \gamma _{\perp
}m_{n\perp }\right) =\int dYG(Y)\prod\nolimits_{k=1}^{n}\int dy_{k}e^{-\beta
\gamma _{\perp }m_{k\perp }\cosh (y_{k}-Y)}.  \label{A}
\end{equation}%
In the above formula the integration over $y_{k}$ is performed over the
specific rapidity bin, depending on an actual measurement. We note that this
bin should be broad enough to \textit{capture} a multi-particle cluster. We
found that $C_{n}$ with $|y_{k}|<2$ and\footnote{%
Of course the broader rapidity bin the better.} $m_{k\perp }>0.6$ GeV can be
very well approximated by $A$ with the integral from $-\infty $ to $+\infty $.\footnote{For smaller $m_{\perp }$ corrections do not exceed $20\%$.} 
In this case $A$ is simply given by a product of $K_{0}$ functions
and we obtain 
\begin{equation}
C_{n}(p_{1\perp },...,p_{n\perp })\sim \int_{1}^{\infty }\frac{d\gamma
_{\perp }}{\gamma _{\perp }^{\kappa }}\prod\nolimits_{k=1}^{n}K_{0}\left(\beta
\gamma _{\perp }m_{k\perp }\right) I_{0}\left( \beta u_{\perp }p_{k\perp }\right) ,
\end{equation}%
which does not depend on the specific shape of $G(Y)$.

\begin{figure}[t]
\begin{center}
\includegraphics[scale=0.37]{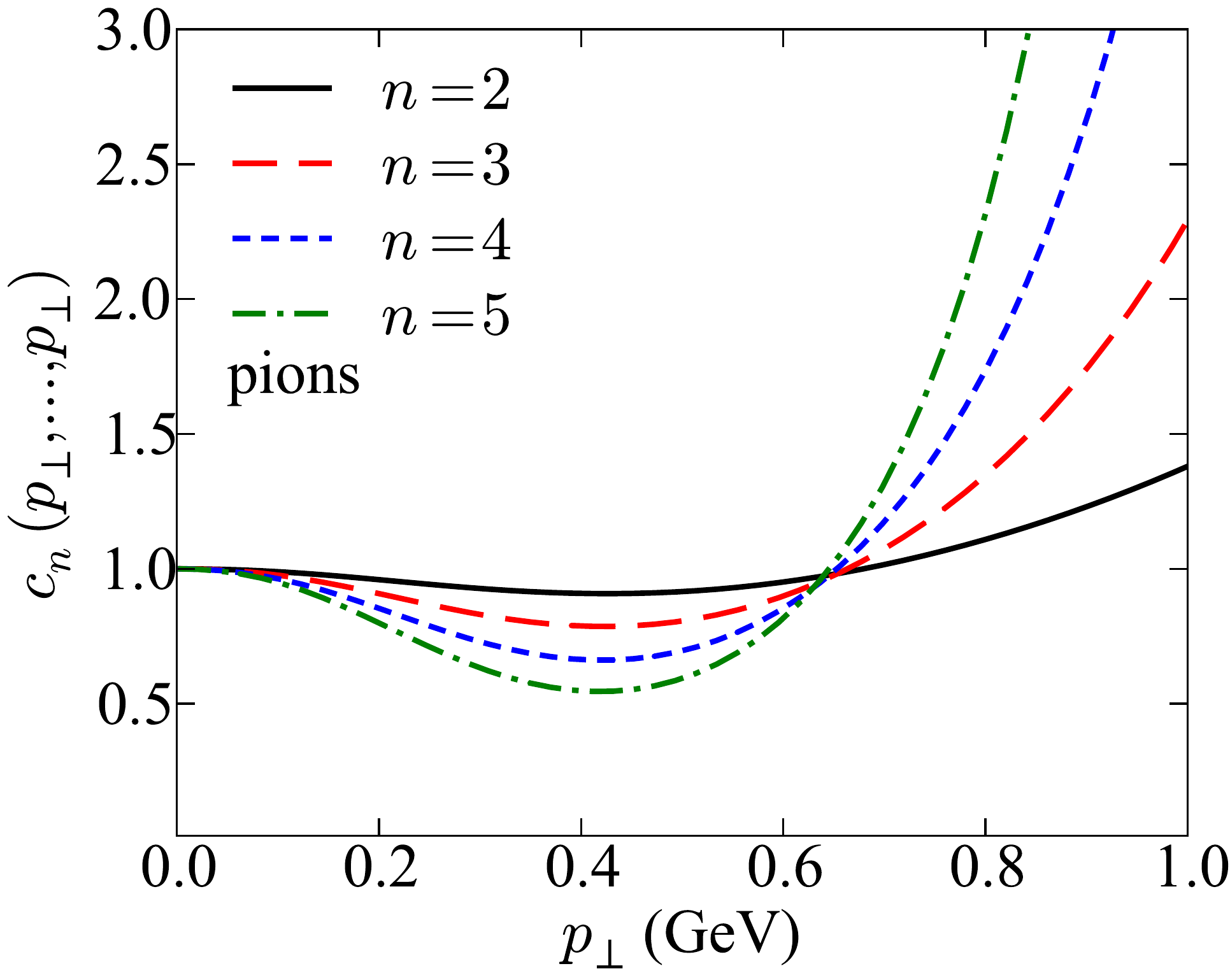}
\includegraphics[scale=0.37]{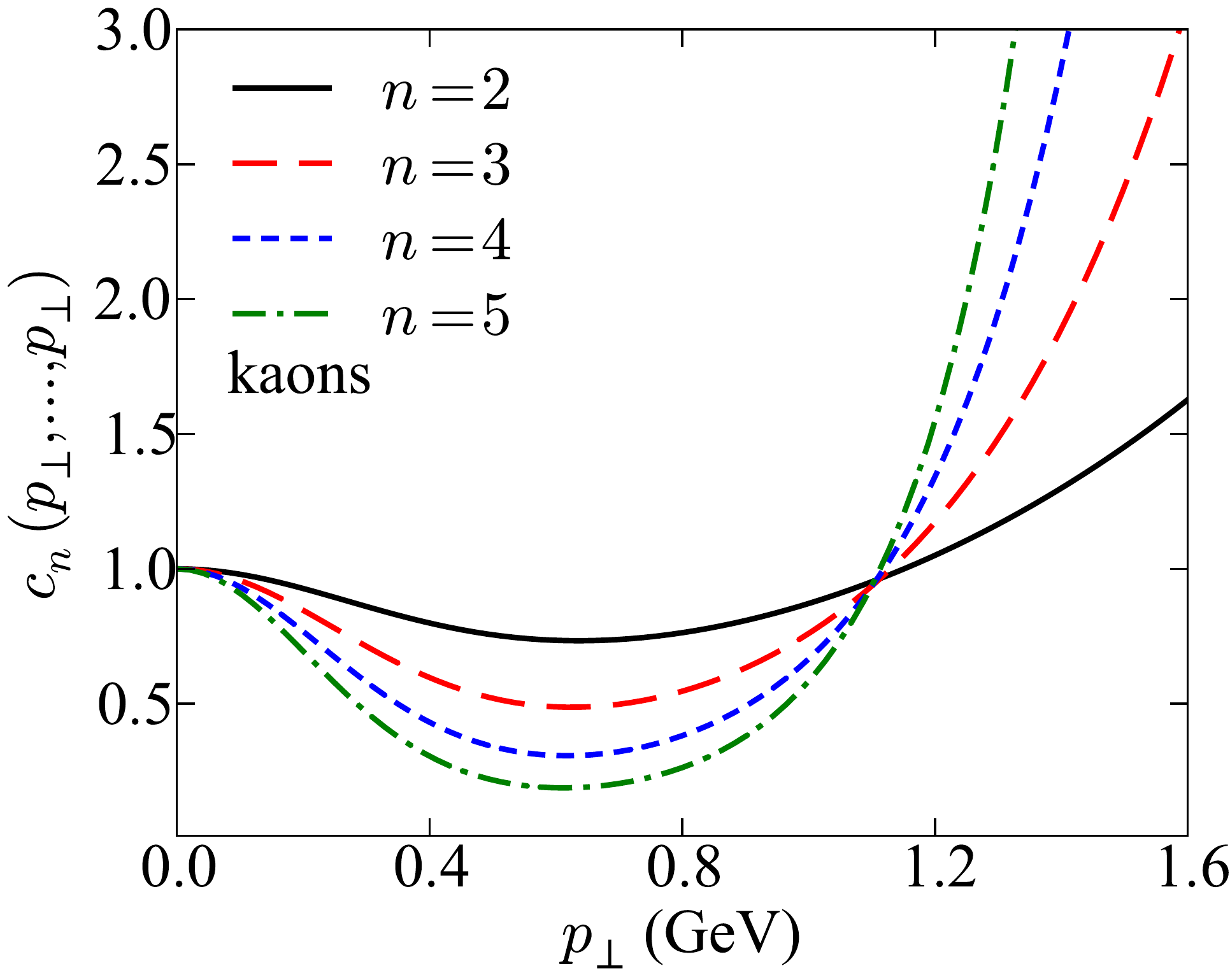}
\end{center}
\par
\vspace{-5mm}
\caption{Multi-particle correlation functions, $c_n(p_{\perp},...,p_{\perp})$, for $n=2,3,4,5$ as a function of $p_{\perp}=p_{1\perp }=...=p_{n\perp}$ for pions (left plot) and kaons (right plot). The curves are scaled to unity at $p_{\perp}=0$.}
\label{fig:1}
\end{figure}

In the following we present our results for 
\begin{equation}
c_{n}(p_{1\perp },...,p_{n\perp })=\frac{C_{n}(p_{1\perp },...,p_{n\perp })}{%
N_{1}(p_{1\perp })\cdots N_{1}(p_{n\perp })}.
\end{equation}

As a reminder, the only two free parameters of the model, $T=140$ MeV and $\kappa=5$, are fixed by fitting (\ref{ik}) to the pion and kaon transverse momentum spectra in p+p collisions at $\sqrt{s}=2.76$ TeV \cite{Bialas:2015oua}. 

In Fig. \ref{fig:1}  $c_{n}(p_{\perp },...,p_{\perp })$, is plotted vs  $p_\perp$ for pions (left plot) and for kaons (right plot). As we are not interested in
the overall normalisation
of $c_{n}$  only the ratio $c_{n}(p_{\perp },...,p_{\perp })/c_{n}(0,...,0)$  is plotted.  We observe a
nontrivial dependence on $p_{\perp }$ which is getting more pronounced with
increasing $n$ and particle mass. This dependence on $p_{\perp}$ is not particularly intuitive since the distribution of particle momenta is determined by a subtle interplay between the  temperature $T$ and the power law $\gamma_\perp$ distribution. However, we checked that if $T$ goes to zero $c_{n}(p_{\perp },...,p_{\perp })$ explodes. This can be easily understood. For $T \rightarrow 0$ particles momenta from a cluster are solely determined by $\gamma_\perp$ and are all identical (in the rest frame of a cluster all momenta are zero). It means that correlation between particles is the strongest possible.

\begin{figure}[t]
\begin{center}
\includegraphics[scale=0.37]{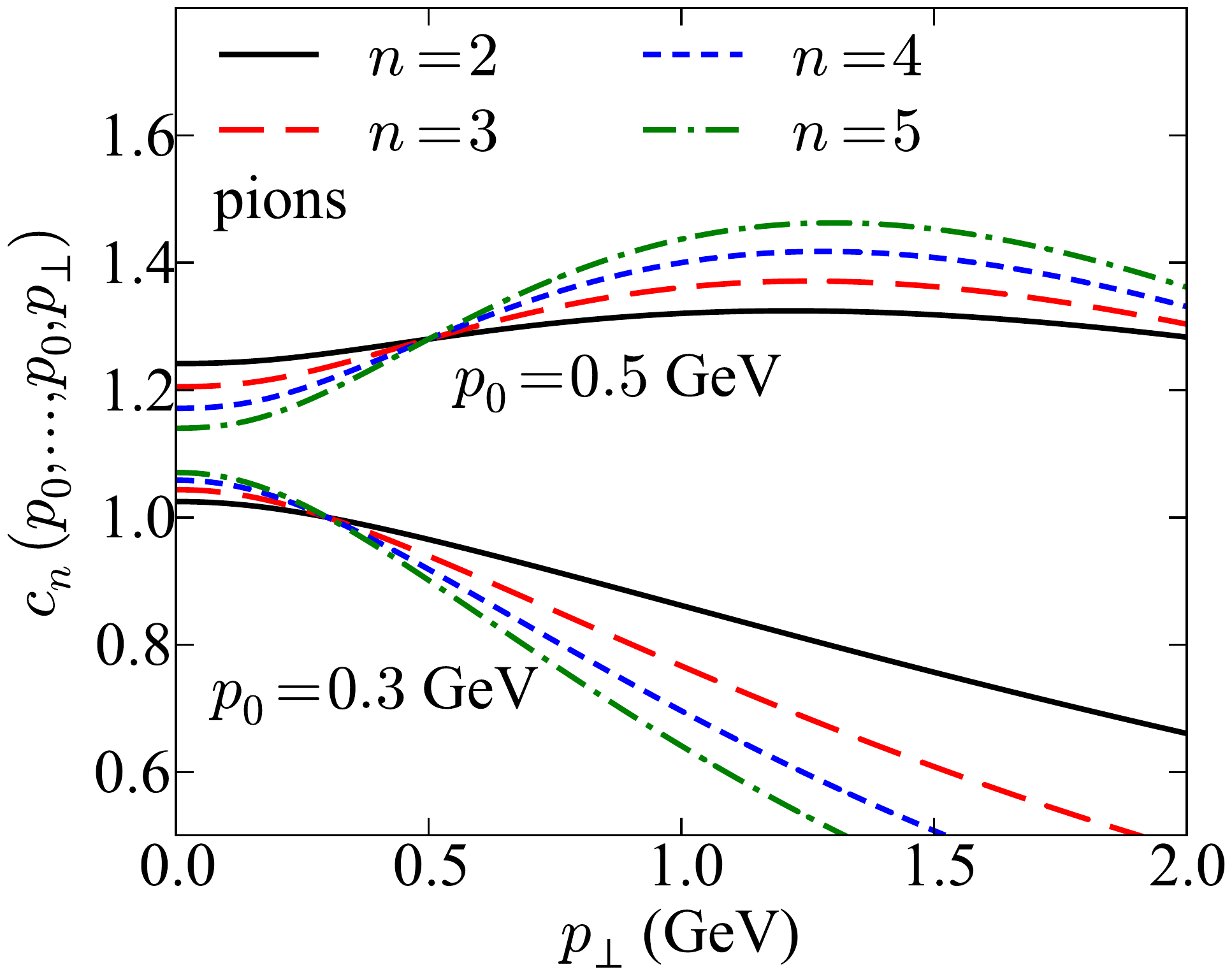}
\includegraphics[scale=0.37]{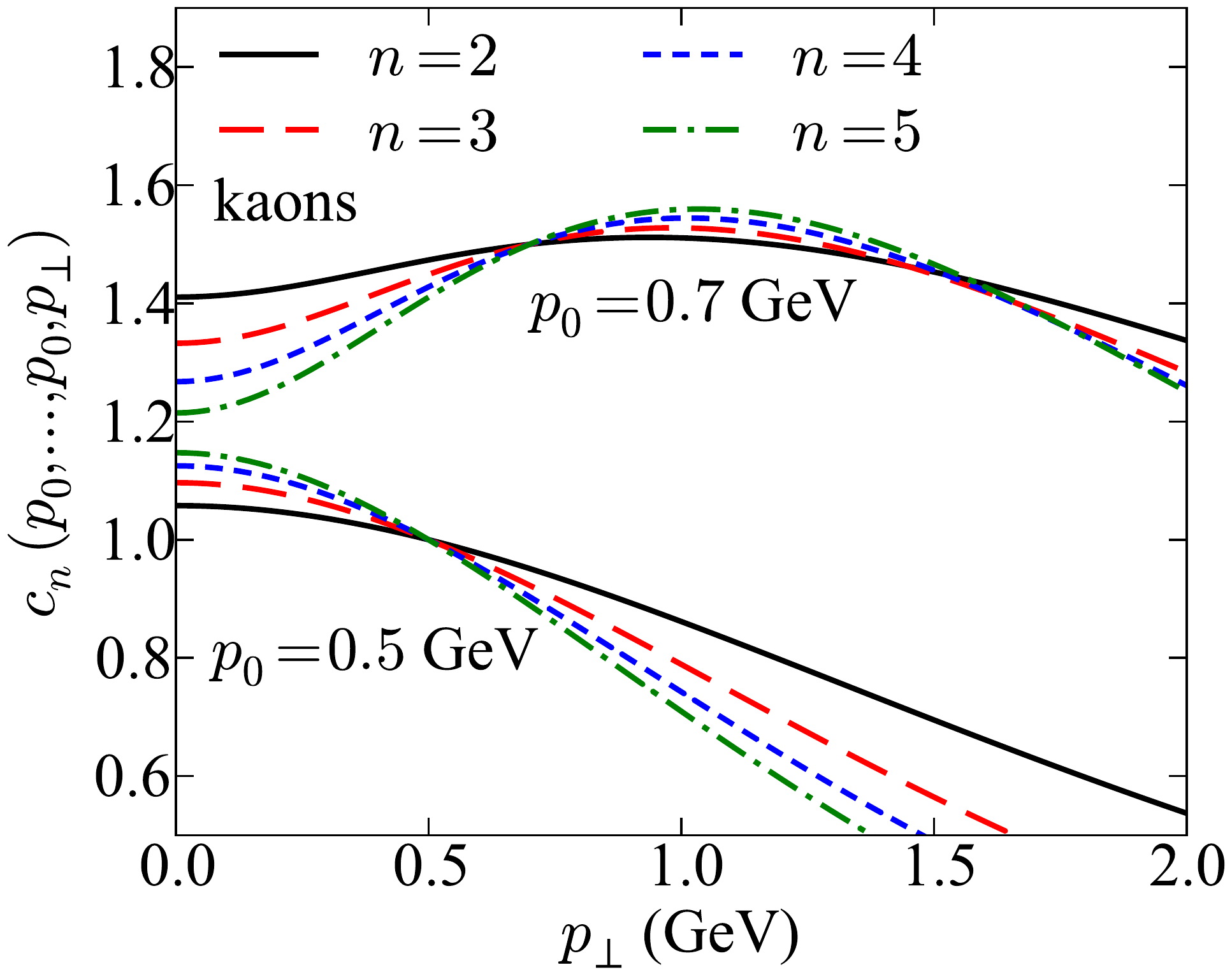}
\end{center}
\par
\vspace{-5mm}
\caption{$n$-particle correlation functions for pions (left) and kaons (right), $c_{n}(p_{0},...,p_{0},p_{\perp })$ for $n=2,3,4,5$ as a function of the transverse momentum of the $n$-th particle, $p_{\perp }$. The remaining $n-1$ particles have fixed momenta $p_{k\perp }=p_{0}$, $k=1,2,...,n-1$. For pions we choose $p_{0}=0.3$ and $0.5$ GeV whereas for kaons we use $p_{0}=0.5$ and $0.7$ GeV. The correlation functions are scaled to unity at $p_{\perp }=p_{0}$. For clarity the curves for $p_{0}=0.5$ GeV (pions) and $p_{0}=0.7$ GeV (kaons) are shifted vertically.}
\label{fig:2}
\end{figure}

In Fig. \ref{fig:2} we study $c_{n}$ for pions (left plot) and kaons (right plot) in a different way. 
We fix $p_{k\perp}=p_{0}$ for $k=1,2,...,n-1$ and study the dependence of 
$c_{n}(p_{0},...,p_{0},p_{n\perp })/ c_{n}(p_{0},...,p_{0},p_{0})$ on the transverse momentum of the last particle
$p_{n\perp}=p_{\perp}$. We choose two values
of $p_{0}=0.3$ and $0.5$ GeV for pions, whereas for kaons we use $p_{0}=0.5$ and $0.7$ GeV.  The results shown  in Fig. \ref{fig:2}, can be easily
understood. When one chooses a certain number of particles from the same cluster
to have  rather small transverse momenta, one selects a cluster with a small transverse velocity and
consequently the last particle is expected to have small $p_{\perp }$ as well. When
 $p_{0}$  increases,  the last particle also tends to have larger $p_{\perp}$. In other words, depending on the value of $p_0$, the correlation function $c_n(p_{0},...,p_{0},p_{\perp })$ can be either decreasing or increasing function of $p_{\perp }$ in the vicinity of $p_0$. However it is somehow surprising that the correlation function is not peaked at $p_{\perp}=p_0$. We verified that this is the consequence of the non-zero temperature.

\begin{figure}[!h]
\begin{center}
\includegraphics[scale=0.37]{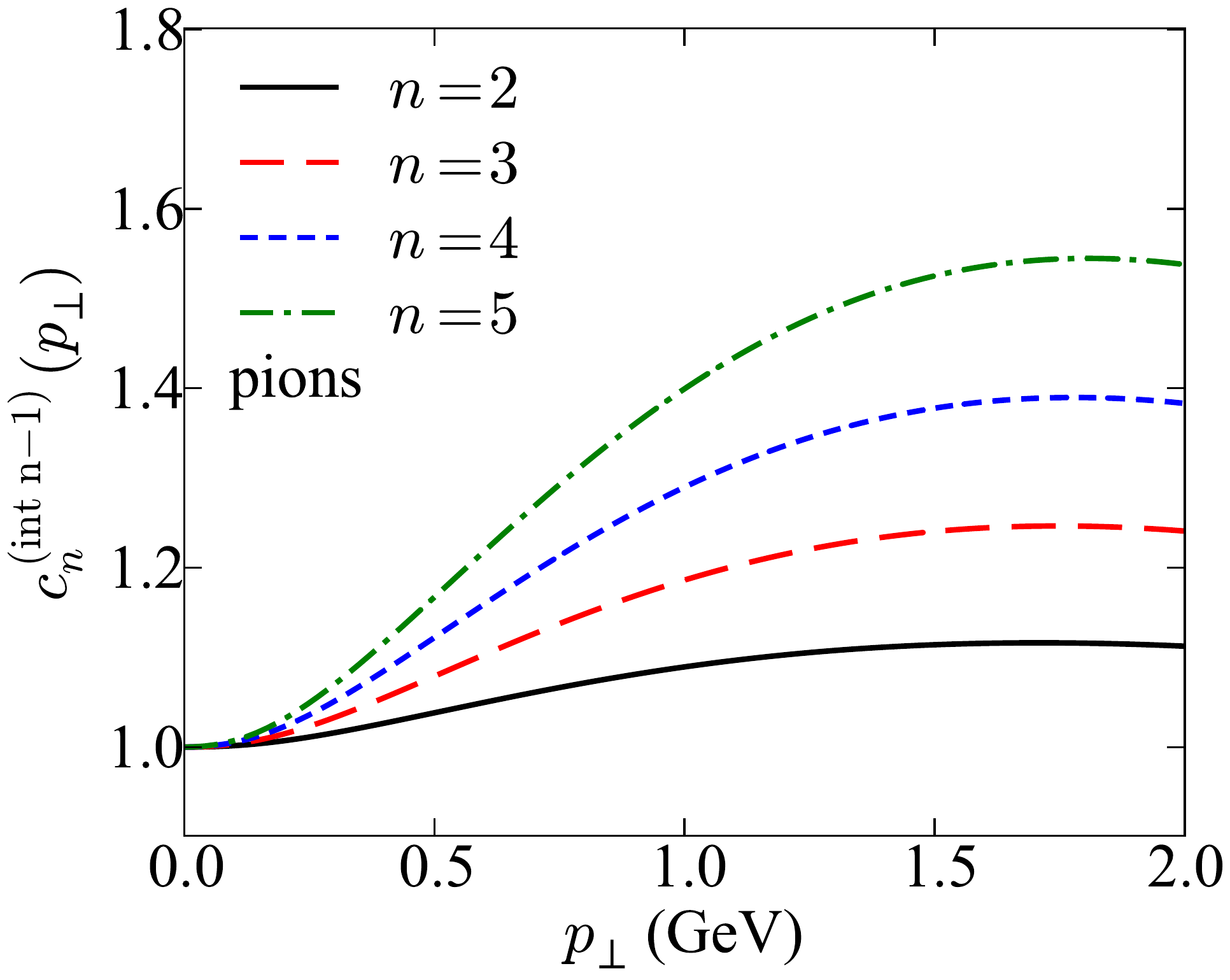}
\includegraphics[scale=0.37]{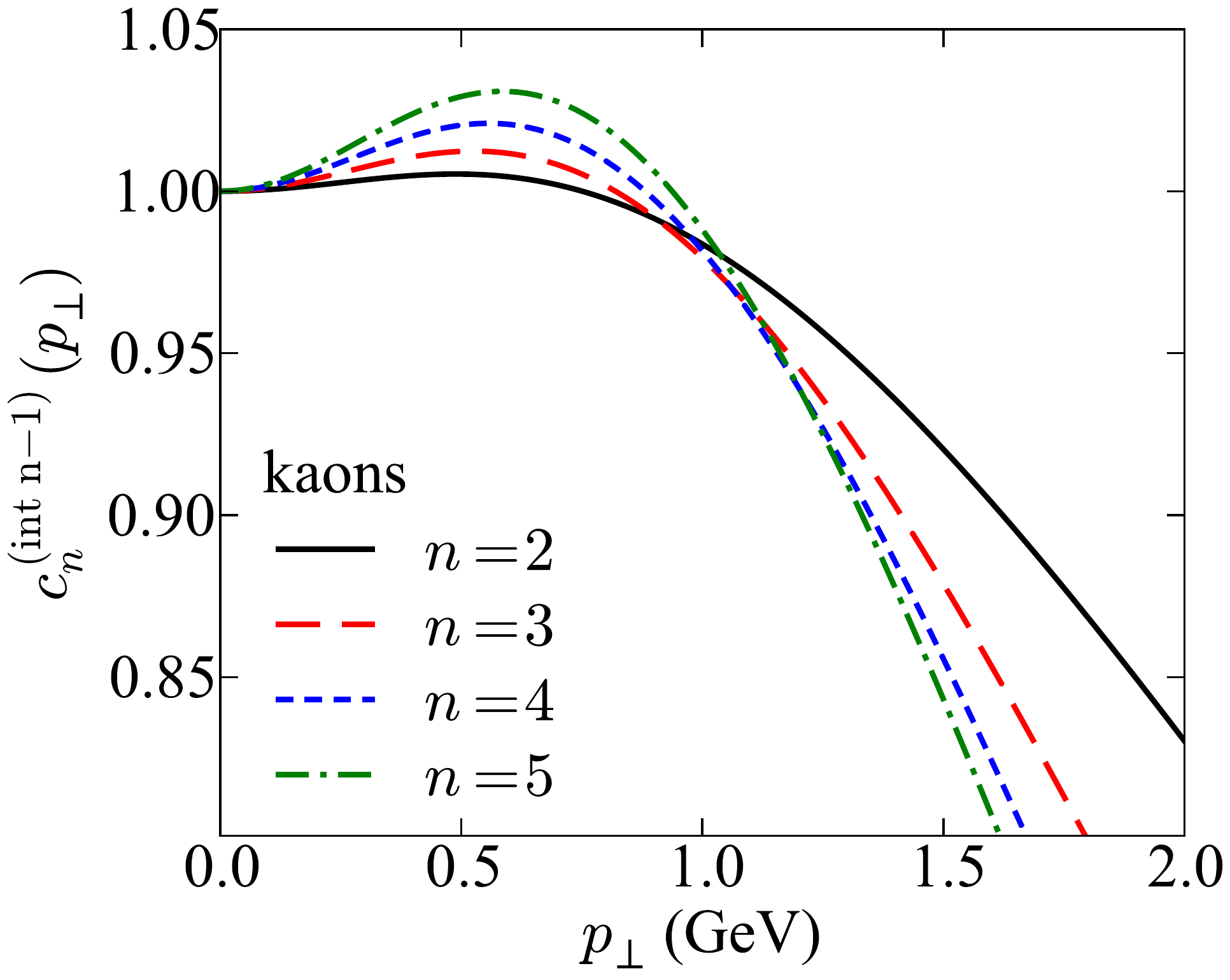}
\end{center}
\par
\vspace{-5mm}
\caption{$n$-particle correlation functions for pions (left) and kaons (right) for integrated $n-1$ transverse momenta in the range  $0.2<p_{k\perp}<1.5$ GeV, $k=2,3,...,n$ as a function of $p_{\perp}$ of the remaining particle. The curves are scaled to unity at $p_{\perp}=0$.}
\label{fig:3}
\end{figure}

Finally, in Fig. \ref{fig:3} we present the results for a more practical situation. We
integrate $n-1$ particles over a  range of transverse momenta and plot $c_n$
as a function of $p_{\perp }$ of the momentum of the remaining particle 
\begin{equation}
c_{n}^{(\text{int }n-1)}(p_{\perp })=\frac{\int dp_{2\perp }^{2}\cdots dp_{n\perp
}^{2}C_{n}(p_{\perp },p_{2\perp },...,p_{n\perp })}{N_{1}(p_{\perp })\int
dp_{2\perp }^{2}N_{1}(p_{2\perp })\cdots \int dp_{n\perp }^{2}N_{1}(p_{n\perp })} .
\end{equation}

In Fig. \ref{fig:3} we show $c_{n}^{(\text{int }n-1)}(p_{\perp })$, $n=2,3,4,5$, where
the transverse momenta, $p_{2\perp },...,p_{n\perp }$, are integrated from $%
0.2$ to $1.5$ GeV. We scale $c_{n}^{(\text{int }n-1)}(p_{\perp })$ to $1$
at $p_{\perp }=0$.

{\bf 3.} To summarise, the formulae for the $n$-particle correlation functions in transverse momenta, following from the production and decay of the uncorrelated statistical clusters, were written down. Numerical evaluation of $n$-particle correlations, $n=2,3,4,5$, were presented and their specific features  discussed. When compared with experimental data, they may serve as a test of the nature of the clustering effects observed since long time in multi-particle production.

Several comments are in order.

(i) The main goal of our investigation is to emphasize the importance of measurements of multi-particle short-range correlations as their presence could be the decisive argument in the discussion on the nature of clustering effects observed in many processes of multi-particles production. We hope that our semi-quantitative estimates shall be useful in the future experimental search.

(ii) To compare our results with experiment it is necessary to measure the
genuine $n$-particle correlation functions, given by the $n$-particle cumulants $C_{n}$ 
(see, e.g., \cite{Botet:2002gj}). For example
\begin{equation}
C_{2}(p_{1},p_{2})=N_{2}(p_{1},p_{2})-N_{1}(p_{1})N_{1}(p_{2}),  \label{C2}
\end{equation}%
and for three particles%
\begin{eqnarray}
C_3(p_1,p_2,p_3)
&=& N_3(p_1,p_2,p_3) +2 N_1(p_1) N_1(p_2) N_1(p_3) - N_2(p_1,p_2)N_1(p_3)- 
\notag \\
&& N_2(p_1,p_3)N_1(p_2)- N_2(p_2,p_3)N_1(p_1), \label{C3} 
\end{eqnarray}
where $N_{i}$ are the standard (inclusive) $i$-particle densities. The expressions for
up to six particles can be found, e.g., in Ref. \cite{Bzdak:2015dja}, see also \cite{Botet:2002gj}. 

(iii) To illustrate the results, we  evaluated numerically multi-particle correlations up to five particles. If needed, it is not difficult to perform similar evaluation for any number of particles.


\bigskip

\noindent{\bf Acknowledgments}
{}\\
This investigation was supported by the Ministry of Science and Higher Education (MNiSW), by founding from the Foundation for Polish Science, and by the National Science Centre (Narodowe Centrum Nauki), Grant Nos. DEC-2013/09/B/ST2/00497 and DEC-2014/15/B/ST2/00175.

\end{document}